\documentclass[a4paper,aps,prd,10pt,preprintnumbers,showpacs,twocolumn,superscriptaddress,nofootinbib,amsmath,amssymb]{revtex4-1}
\usepackage[dvips]{graphics}

\begin{document}
\title{No stable wormholes in Einstein-dilaton-Gauss-Bonnet theory}
\author{M. A. Cuyubamba}\email{marco.espinoza@ufabc.edu.br}
\affiliation{Centro de Matem\'atica, Computa\c{c}\~ao e Cogni\c{c}\~ao (CMCC), Universidade Federal do ABC (UFABC),\\ Rua Aboli\c{c}\~ao, CEP: 09210-180, Santo Andr\'e, SP, Brazil}
\author{R. A. Konoplya}\email{konoplya\_roma@yahoo.com}
\affiliation{Theoretical Astrophysics, Eberhard-Karls University of T\"ubingen, T\"ubingen 72076, Germany}
\affiliation{Peoples Friendship University of Russia (RUDN University), 6 Miklukho-Maklaya Street, Moscow 117198, Russian Federation}
\affiliation{Institute of Physics and Research Centre of Theoretical Physics and Astrophysics, Faculty of Philosophy and Science, Silesian University in Opava, CZ-746 01 Opava, Czech Republic}
\author{A. Zhidenko}\email{olexandr.zhydenko@ufabc.edu.br}
\affiliation{Centro de Matem\'atica, Computa\c{c}\~ao e Cogni\c{c}\~ao (CMCC), Universidade Federal do ABC (UFABC),\\ Rua Aboli\c{c}\~ao, CEP: 09210-180, Santo Andr\'e, SP, Brazil}

\begin{abstract}
In \cite{Kanti:2011jz} 
it was shown that the four-dimensional Einstein-dilaton-Gauss-Bonnet theory allows for wormholes without introducing any exotic matter. The numerical solution for the wormhole was obtained there and it was claimed that this solution is gravitationally stable against radial perturbations, what, by now, would mean the only known theoretical possibility for existence of an apparently stable, four-dimensional and asymptotically flat wormhole without exotic matter. Here, more detailed analysis of perturbations shows that the Kanti-Kleihaus-Kunz wormhole is unstable against small perturbations for any values of its parameters. The exponential growth appears in the time domain after a long period of damped oscillations, in the same way as it takes place in the case of unstable higher-dimensional black holes in the Einstein-Gauss-Bonnet theory. The instability is driven by the purely imaginary mode, which is \emph{nonperturbative} in the Gauss-Bonnet coupling $\alpha$.
\end{abstract}
\maketitle

\section{Introduction}

Wormholes are so far theoretical objects linking different points in spacetime or even different universes. A number of recent works are devoted to potentially observable features of wormholes, such as gravitational lensing \cite{Tsukamoto:2016qro,Tsukamoto:2017edq}, quasinormal modes \cite{Konoplya:2005et,Konoplya:2010kv,Aneesh:2018hlp}, accretion \cite{Zhou:2016koy} and others \cite{Bueno:2017hyj,Nascimento:2017def,Nandi:2016uzg}. A wormhole can mimic the gravitational wave response of a black hole to the external perturbation  both at the intermediate \cite{Damour:2007ap} and late \cite{Konoplya:2016hmd} times.

Usually for a wormhole to be traversable an \emph{exotic matter} with negative energy density is necessary, in order to create strong repulsive gravitational force preventing the wormhole's throat from shrinking. A fortunate exception was suggested in \cite{Kanti:2011jz}, where the numerical solution for a four-dimensional asymptotically flat wormhole was found in the Einstein-dilaton-Gauss-Bonnet theory.

This theory represents string theory inspired corrections to the Einstein theory at low energies. The full heterotic string effective theory includes also axions, fermions and gauge fields as well as higher than the second order curvature corrections \cite{Gross:1986mw,Metsaev:1987zx}. In this framework the Einstein-dilaton-Gauss-Bonnet action corresponds to a kind of minimal effective theory. Therefore, there is a great interest in understanding physics of various compact objects, first of all, such as black holes and neutron stars, in the Einstein-dilaton-Gauss-Bonnet theory \cite{Kokkotas:2017ymc,Antoniou:2017hxj,Hartmann:2013tca,Blazquez-Salcedo:2015ets,Doneva:2017bvd,Silva:2017uqg,Zhang:2017unx,Gwak:2017fea,Blazquez-Salcedo:2016yka,Antoniou:2017acq,Nampalliwar:2018iru}. The negative energy density, which is necessary for existence of a wormhole, is provided by the Gauss-Bonnet term. Thus, the solution suggested in \cite{Kanti:2011jz} is an important example of a wormhole without any exotic matter, which is stipulated by the fundamental physics.

An essential criterium related to theoretical possibility of existence of wormholes is their stability against small spacetime perturbations. Stability of wormholes was studied in a number of works \cite{Poisson:1995sv,Lobo:2005yv,Eiroa:2003wp,Dzhunushaliev:2017syc}. Stability of thin-shell wormholes was investigated only against most-stable purely radial perturbations \cite{Poisson:1995sv,Lobo:2005yv,Eiroa:2003wp}, while (also radial) stability of wormholes in general relativity reported in \cite{Bronnikov:2013coa} requires a rather odd equation of state for the surrounding matter \cite{Konoplya:2016hmd}. Various wormholes with ghost scalar field are known to be unstable \cite{Gonzalez:2008wd,Bronnikov:2012ch}. Thus, to the best of our knowledge there is no example of a four-dimensional asymptotically flat wormhole solution whose stability would be well established.

Therefore, the stability of such a wormhole solution against small spacetime perturbations is important to prove the viability of the wormhole model. Spherically symmetric perturbations were considered in  \cite{Kanti:2011jz} and there it was concluded that the wormhole is stable against spherical perturbations. The boundary conditions used in \cite{Kanti:2011jz} imply fixing the size of the wormhole throat, which looks nonphysical.  This was motivated in \cite{Kanti:2011jz} by the requirement that small perturbation of the dilaton $\delta \phi$ remains finite in the fixed point of the wormholes' throat, and the Dirichlet boundary condition was imposed there. This effectively disconnected the two regions to the left and right from the throat.

Here we show that, if one relaxes the above requirement by allowing for the perturbations of the throat size, the correct boundary conditions must be the same as for an asymptotically flat black hole: purely outgoing waves at both spacial asymptotical regions. Solving the regularized wave equations under these boundary conditions leads to instability of traversable wormholes in the Einstein-dilaton-Gauss-Bonnet theory at whatever small values of the coupling constants.

The paper is organized as follows. In Sec.~\ref{sec:sol} we reobtain the numerical wormhole metric of \cite{Kanti:2011jz} and discuss the allowed range of parameters and basic features of the solution. Sec.~\ref{sec:perturb} deals with the perturbation equations. In  Sec.~\ref{sec:res} we discuss the results of the time-domain integration and the found instability. Finally, in Sec.~\ref{sec:final} we explain the reason for the discrepancy between our work and \cite{Kanti:2011jz}, and summarize the obtained results.

\section{The Kanti-Kleihaus-Kunz wormhole solution}\label{sec:sol}

The low-energy heterotic string theory is described by the action
\cite{Gross:1986mw,Metsaev:1987zx}
\begin{eqnarray}\label{act}
S&=&\frac{1}{16 \pi}\int d^4x \sqrt{-g} \Biggr[R - \frac{1}{2}
 \partial_\mu \phi \,\partial^\mu \phi
+ \alpha  e^{-\gamma \phi} R^2_{\rm GB}	  \Biggr],
\end{eqnarray}
where $\alpha$ is a positive parameter proportional to the Regge slope, $\gamma$ is the dilaton coupling constant and
$$R^2_{\rm GB} = R_{\mu\nu\rho\sigma} R^{\mu\nu\rho\sigma} - 4 R_{\mu\nu} R^{\mu\nu} + R^2.$$

The equations of motion for the dilaton and gravitational fields are given by
\begin{eqnarray}
&\nabla^2 \phi & = \alpha \gamma  e^{-\gamma \phi}R^2_{\rm GB},	
\label{eqs}\\
& G_{\mu\nu} & =
\frac{1}{2}\left[\nabla_\mu \phi \nabla_\nu \phi
                 -\frac{1}{2}g_{\mu\nu}\nabla_\lambda \phi \nabla^\lambda\phi
		 \right]
\nonumber\\
& &
-\alpha e^{-\gamma \phi}
\left[	H_{\mu\nu}
  +4\left(\gamma^2\nabla^\rho \phi \nabla^\sigma \phi
           -\gamma \nabla^\rho\nabla^\sigma \phi\right)	P_{\mu\rho\nu\sigma}
		 \right],
\nonumber
\end{eqnarray}
where
\begin{eqnarray}\nonumber
H_{\mu\nu} & = & 2\left[R R_{\mu\nu} -2 R_{\mu\rho}R^\rho_\nu
                        -2 R_{\mu\rho\nu\sigma}R^{\rho\sigma}
		   \right]
\\\nonumber && +2R_{\mu\rho\sigma\lambda}R_\nu^{\ \rho\sigma\lambda}-\frac{1}{2}g_{\mu\nu}R^2_{\rm GB}	\ ,
\\\nonumber
P_{\mu\nu\rho\sigma} & = & R_{\mu\nu\rho\sigma}
+2 g_{\mu\sigma} R_{\rho\nu}-2 g_{\mu\rho} R_{\sigma\nu}
\\\nonumber &&
+2 g_{\nu\rho} R_{\sigma\mu}-2 g_{\nu\sigma} R_{\rho\mu}
+R g_{\mu\rho} g_{\sigma\nu}-R g_{\mu\sigma} g_{\rho\nu} \ .
\label{HP}
\end{eqnarray}

The above equations allow for asymptotically flat spherically symmetric wormholes that can be described by the following line element \cite{Kanti:2011jz}
\begin{eqnarray}
ds^2 &=& -e^{2\nu(l)}dt^2+f(l)dl^2
+r^2\left(d\theta^2+\sin^2\theta d\varphi^2 \right)\ ,\nonumber
\\ r^2&\equiv& l^2+r_0^2\ ,\label{metricL}
\end{eqnarray}
where $r_0$ is the radius of the throat. The metric functions, $f(l)$ and $\nu(l)$, and the dilaton field $\phi(l)$ satisfy the equations \cite{Kanti:2011yv}
\begin{widetext}
\begin{eqnarray}
f' + \frac{f (r^2 f + l^2 - 2 r^2)}{l r^2}
& = &
\frac{r^2 f \phi'^2}{4 l}
 +2\alpha \gamma\frac{e^{-\gamma\phi}}{l r^2} \Biggr\{
  2 (r^2 f - l^2)(\gamma\phi'^2-\phi'')
 + \phi'\left[\frac{f'}{f}\,(r^2 f-3 l^2) +\frac{4 l r_0^2}{r^2}\right]\Biggr\},  \nonumber\\
\nu' - \frac{r^2 f -l^2}{2 l r^2}
& = &
\frac{\phi'^2 r^2}{8 l}
+2\alpha \gamma\frac{e^{-\gamma\phi}}{l r^2 f}
\,\nu'\phi' (r^2 f- 3 l^2)\,,\label{ode}
\\
\phi'' + \nu'\phi' + \frac{\phi' (4 l f - r^2 f')}{2 r^2 f}
& = &
4\alpha \gamma\frac{e^{-\gamma\phi}}{r^4 f} \Biggr\{
-2 (r^2 f - l^2)(\nu'^2+\nu'')
+\nu'\left[\frac{f'}{f}\,(r^2 f-3 l^2) +\frac{4 l r_0^2}{r^2}\right]\Biggr\},
\nonumber
\end{eqnarray}
\end{widetext}
where the prime designates the derivative with respect to a radial coordinate $l$.

\begin{figure*}
\resizebox{\linewidth}{!}{\includegraphics*{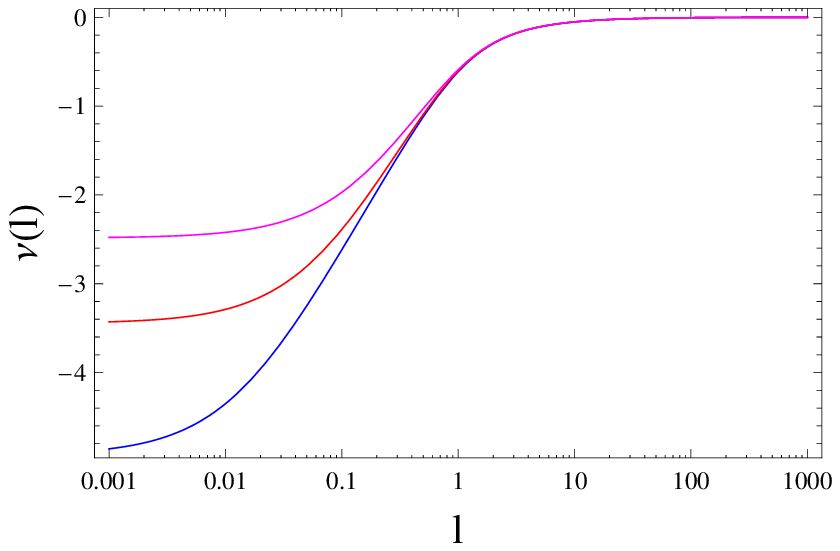}\includegraphics*{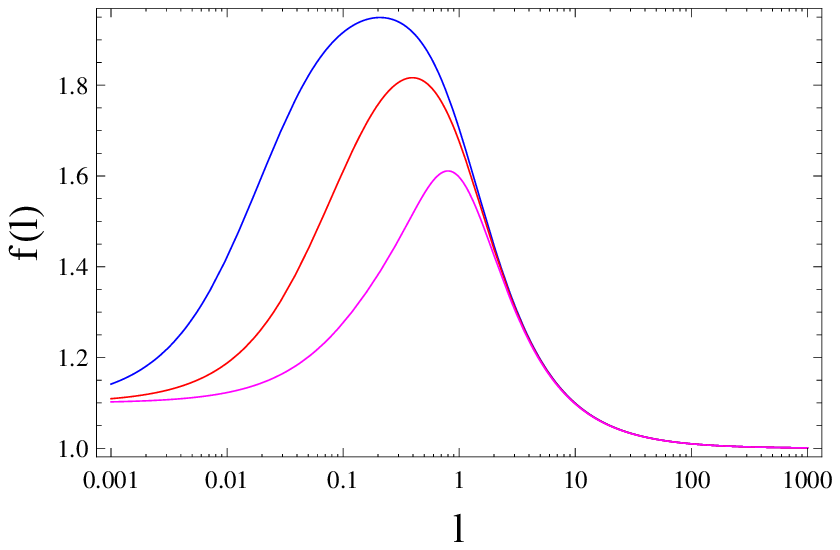}}
\caption{Metric functions, $\nu(l)$ (left panel, from bottom to top) and $f(l)$ (right panel, from top to bottom), for the wormholes $\alpha=0.005r_0^2$ (blue), $\alpha=0.02r_0^2$ (red), and $\alpha=0.05r_0^2$ (magenta) for $f_0=1.1$.}\label{mplot}
\end{figure*}

The following initial conditions are imposed at the throat ($l=0$)
\begin{eqnarray}
\nonumber f(0)&=&f_0,\\
\nonumber \phi(0)&=&\phi_0,\\
\nonumber \nu(0)&=&\nu_0,\\
\nonumber \phi'(0)^2 &=& \frac{f_0(f_0-1)}{2\alpha\gamma^2 e^{-\gamma \phi_0}
\left[f_0-2(f_0-1)\frac{\alpha}{r_0^2} e^{-\gamma \phi_0}\right]}\ .
\end{eqnarray}
The latter of the four conditions follows from the requirement of regularity of the metric coefficients at the throat. The requirement
\begin{equation}
f(l \to \infty) \to 1
\end{equation}
following from asymptotic flatness is always satisfied.
Due to the scaling symmetry (see Sec. IIC in \cite{Kanti:2011yv}), without loss of generality we take $\gamma=1$ and $r_0=1$, i.~e. we measure $\alpha$ and all dimensional quantities in the units of $r_0$. Then, for given $\alpha>0$ and $f_0>1$, we choose $\nu_0$ and $\phi_0$ such that
\begin{equation}
\lim_{l\to \infty}\nu = 0 \,, \qquad \lim_{l\to \infty}\phi = 0 \, .
\end{equation}
Once the units are chosen as above, the wormhole solution can be found for every value of $\alpha$ for which (see Fig.~\ref{mplot})
\begin{equation}
\frac{\alpha}{r_{0}^2}\lessapprox 0.13.
\end{equation}

The above setup defines two families of the solutions to the equations (\ref{ode}) for $l\geq0$ corresponding to two possible signs of $\phi'(0)$. All the wormhole configurations presented in \cite{Kanti:2011yv} correspond to the choice $\phi'(0)<0$. It is clear that if one requires the smoothness of the solution at $l=0$ the resulting configuration is not symmetric with respect to the throat. Indeed, if one replaces $l\to-l$, then $\phi'(0)\to-\phi'(0)$ (as well as derivatives of the metric functions). It turns out that such solutions have singularities after crossing the throat. Therefore, we shall study symmetric wormholes such that $\phi'(+0)=-\phi'(-0)<0$, i.~e. with discontinuities of the first derivatives at the throat. Although such a geometry looks artificial, it apparently does not lead to problems because the observable quantities remain finite at the throat \cite{Kanti:2011yv}. The discontinuities can be attributed to the presence of matter with positive energy density and pressure at the throat \cite{Kanti:2011jz}.

\section{Linearized spherically symmetric perturbations}\label{sec:perturb}

The equations for the linearized spherically symmetric perturbations of the dilaton $\delta\phi$ and the metric functions, $\delta f$ and $\delta\nu$,
\begin{eqnarray}\nonumber
\phi\to\phi(l)+\delta\phi(t,l)\\\label{perturbations}
f\to f(l)+\delta f(t,l)\\
\nu\to\nu(l)+\delta\nu(t,l)\nonumber
\end{eqnarray}
can be reduced to one dynamical equation for $\delta\phi$ \cite{Kanti:2011yv}
\begin{equation}
\frac{\partial^2\delta\phi}{\partial l^2} + q_1(l) \frac{\partial\delta\phi}{\partial l} + q_0(l) \delta\phi(t,l) - q_\sigma(l) \frac{\partial^2\delta\phi}{\partial t^2} = 0 \ ,
\label{lineq}
\end{equation}
while the perturbations of the metric functions, $\delta f$ and $\delta\nu$, can be expressed in terms of $\delta\phi$ and its derivatives. Notice also that the perturbation of the throat radius $\delta r(l, t)$ is set to $0$ in \cite{Kanti:2011yv}.

The coefficients $q_\sigma(l)$, $q_0(l)$ and $q_1(l)$ can be expressed in terms of the background solution and have quite a cumbersome form. It was found in \cite{Kanti:2011yv} that, for small wormholes ($\alpha>0.05r_0^2$), $q_\sigma(l)<0$ for some interval of $l$, implying that the kinetic term has a wrong sign. A similar feature was observed for the small black holes in the Lovelock theory \cite{Takahashi:2010gz}. Such black holes are always linearly unstable with respect to the gravitational perturbations of the vector type \cite{Konoplya:2017lhs}. Yet, it was pointed out in \cite{Reall:2014pwa} that the nonhyperbolicity of the perturbation equations implies that the
stability of such black holes is not a well-posed problem. In the present paper we do not consider the wormholes for which perturbation equations are nonhyperbolic.

For sufficiently large wormholes ($\alpha\leq0.05r_0^2$) the function $q_\sigma(l)$ is positive everywhere and (\ref{lineq}) can be reduced to the wavelike equation \cite{Kanti:2011yv}
\begin{equation}
\frac{\partial^2\chi}{\partial t^2}-\frac{\partial^2\chi}{\partial y^2} + V_{\rm eff}(l)\chi(t,l)= 0 \ ,
\label{wavelike}
\end{equation}
with the effective potential
\begin{equation}\label{potential}
V_{\rm eff}(l)=\frac{q_0(l)}{q_\sigma(l)}-\frac{q_1'(l)}{2 q_\sigma(l)}-\frac{q_1(l)^2}{4 q_\sigma(l)}-\frac{q_\sigma''(l)}{4q_\sigma(l)^2}+\frac{5 q_\sigma'(l)^2}{16q_\sigma(l)^3},
\end{equation}
where $y$ is the tortoise coordinate defined as
$$y= \int\sqrt{q_\sigma(l)}dl.$$

Although $q_0(l)$ and $q_1(l)$ diverge at the throat, the effective potential (\ref{potential}) is finite everywhere. In order to find quasinormal modes for the above case, we need to impose purely outgoing wave boundary conditions at both infinities.

\begin{equation}
\chi  \sim e^{\pm i \omega y}, \quad y \to \pm \infty
\end{equation}

This is compatible with the finite effective potential which approaches constant values at both asymptotic regions \cite{Konoplya:2005et}. In a similar fashion with the black hole case (see, for example, \cite{Konoplya:2011qq}), such boundary conditions describe the ``momentary'' reaction of a wormhole to the perturbation, when the source of perturbation stopped acting. Therefore, in the next section we use the method of numerical integration of the wavelike equation \cite{Gundlach:1993tp} which was previously used mostly for black holes.

\begin{figure*}
\resizebox{\linewidth}{!}{\includegraphics*{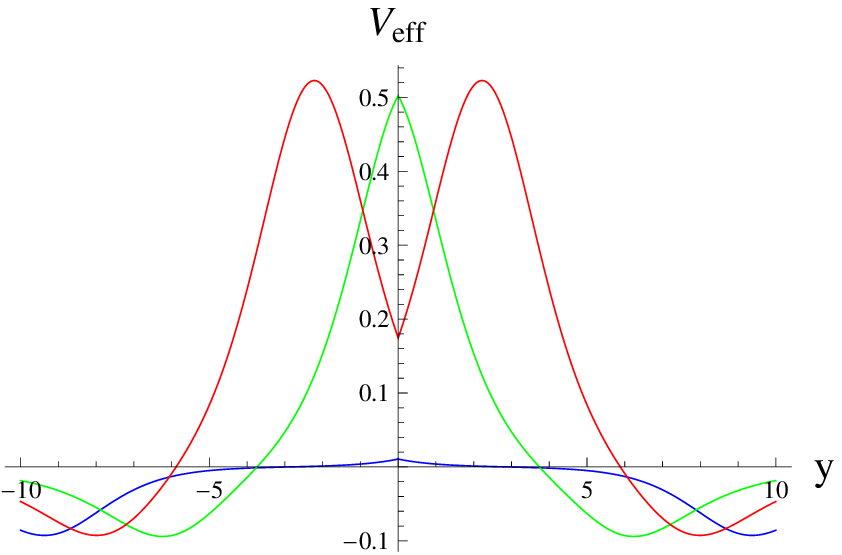}\includegraphics*{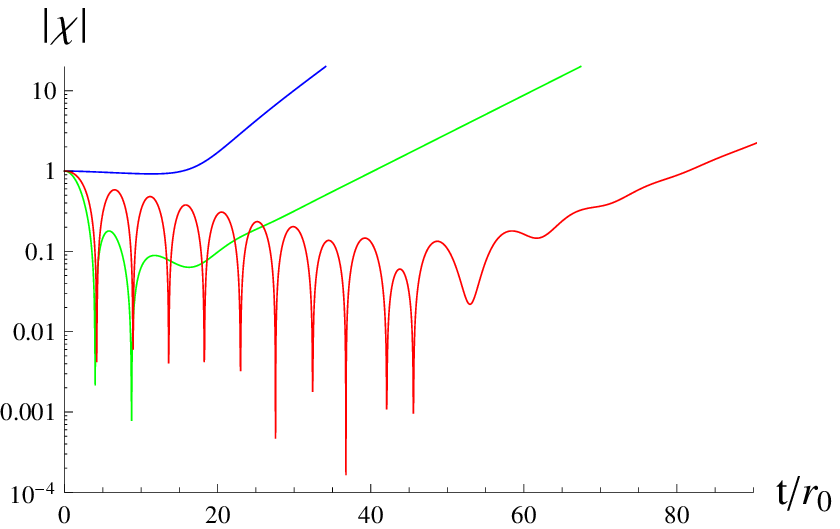}}
\caption{Effective potentials (left panel) and the time-domain profiles (right panel) for the spherically symmetric perturbations of the wormhole $\alpha=0.02r_0^2$: $f_0=1.001$ (blue), $f_0=1.1$ (green), $f_0=10$ (red). As $f_0$ grows the peak of the potential becomes larger at the throat and eventually shifts outside the throat, leading to duplication of the peak due to symmetry. The unstable mode appears later in the profile and grows slower. }\label{aplot}
\end{figure*}

\begin{figure*}
\resizebox{\linewidth}{!}{\includegraphics*{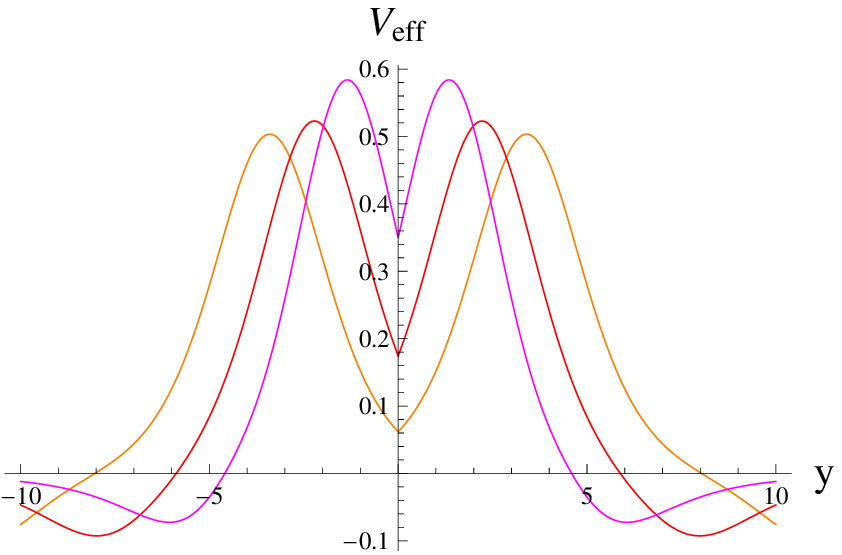}\includegraphics*{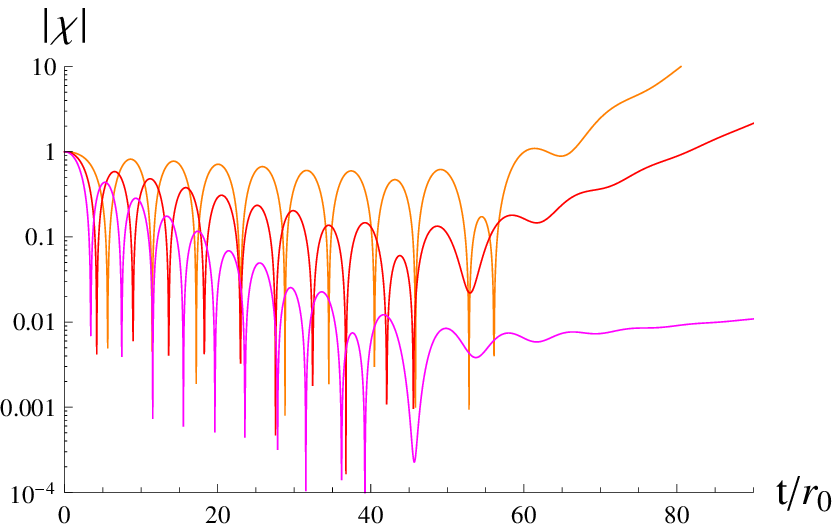}}
\caption{Effective potentials (left panel) and the time-domain profiles (right panel) for the spherically symmetric perturbations of the wormhole $f_0=10$: $\alpha=0.005r_0^2$ (orange), $\alpha=0.02r_0^2$ (red), $\alpha=0.05r_0^2$ (magenta). As $\alpha$ grows the peak becomes higher and shifts towards the throat uppering the local minimum at the throat. The perturbations are unstable because the effective potentials remain negative dominant.}\label{fplot}
\end{figure*}

\section{Time domain profiles and the instability}\label{sec:res}

Following \cite{Bronnikov:2011if}, for the general spherically symmetric perturbations we study the time-domain profile for $\chi(t,l)$ at the throat ($l=0$).  In order to obtain the time-domain profile we use the discretization scheme proposed by Gundlach, Price, and Pullin \cite{Gundlach:1993tp}. Rewriting (\ref{wavelike}) in terms of the light-cone coordinates $du=dt+dy$ and $dv=dt-dy$, one finds
\begin{equation}\label{timedomain}
4\frac{\partial^2\chi}{\partial u\partial v}=-V_{\rm eff}\left(\frac{u-v}{2}\right)
\chi.
\end{equation}

The discretization scheme has the following form
\begin{eqnarray}\label{ci1}
\chi(N)&=&\chi(W)+\chi(E)-\chi(S)\\\nonumber&&-\frac{\Delta^2}{8}V_{\rm eff}(S)\left[\chi(W)+\chi(E)\right]+\mathcal{O}(\Delta^4),
\end{eqnarray}
where $N$, $M$, $E$, and $S$ are the points of a square in a grid with step $\Delta$ in the discretized $u$-$v$ plane: $S=(u,v)$, $W=(u+\Delta,v)$, $E=(u,v+\Delta)$, and $N=(u+\Delta,v+\Delta)$. With the initial data specified on two null surfaces $u = u_0$ and $v = v_0$, we are able to find values of the function $\chi$ at each of the points of the grid (see \cite{Konoplya:2011qq} for more details). The resulting profiles are qualitatively similar for Gaussian waves and constant initial values.

At the early stage of the quasinormal ringing we observe decaying modes that give way to a nonoscillating growth at late times (see Figs.~\ref{aplot}~and~\ref{fplot}). One could think that such a behavior is the result of some numerical error accumulated when integrating until very late time. However, increasing the accuracy goal of our calculations, as well as diminishing the size of the grid, allows us to achieve the convergence of the whole procedure of the numerical integration. After all, the same behavior (i.e. exponential growth after a long period of damped oscillations) was observed when considering perturbations of the higher-dimensional black holes in the Einstein-Gauss-Bonnet theory \cite{Konoplya:2008ix,Cuyubamba:2016cug}. There the instability of black holes was developing at high values of the multipole number $\ell$ and, therefore, was called \emph{the eikonal instability} \cite{Konoplya:2017ymp}. The parametric region of this instability was coinciding with the region of nonhyperbolicity of the wave equation \cite{Reall:2014pwa}, in which case no good initial value problem can be posed. Thus, this instability is more than a usual instability and implies that its onset breaks down even the concept of small perturbations.

Having the above similarities of the time-domain evolutions for black holes and wormholes in Gauss-Bonnet theories in mind, it would be interesting to test the hyperbolicity of the wave equations for the Einstein-dilaton-Gauss-Bonnet wormholes, as it is not excluded that the instability observed here might be somehow correlated with the nonhyperbolicity.

In Fig.~\ref{fplot} one can see that the smaller values of $\alpha$ correspond to the higher growth rates, so that when $\alpha \to 0$ it is possible that the instability growth rate may go to infinity. This means that the purely imaginary quasinormal frequency, which is responsible for the instability would increase unboundedly as $\alpha$ goes to zero. This indicates that the above mode is \emph{nonperturbative in $\alpha$}. The wormhole is therefore evidently unstable at whatever small value of $\alpha$. The same phenomenon takes place for $D>4$ dimensional asymptotically flat and anti-de Sitter black holes in various theories with higher curvature corrections
\cite{Konoplya:2008ix,Cuyubamba:2016cug,Konoplya:2017ymp,Gonzalez:2017gwa,Grozdanov:2016vgg,Grozdanov:2016fkt,Konoplya:2017zwo}.

\section{Final remarks}\label{sec:final}

However unstable, the wormhole solution obtained in \cite{Kanti:2011jz} is an important example of a wormhole supported by the string theory inspired modification of Einstein equations rather than by some exotic matter. Let us discuss here the reason for the discrepancy between our instability and the stability reported in \cite{Kanti:2011jz}.

The perturbation function in \cite{Kanti:2011jz} is
$$\delta\phi(t,l)=A(l)\chi(t,l).$$
Since the factor $A(l)$ diverges at the throat as ${\cal O}(l^{-1})$, the vanishing boundary conditions were imposed in \cite{Kanti:2011yv} in order to have finite perturbations at the throat. Effectively this disconnects the regions of space on both sides of the throat for purely radial modes of perturbation.

It is important to notice that $\delta\phi$ is not a gauge invariant quantity. Equation (\ref{lineq}) was derived assuming that $\delta r=0$ in
\begin{equation}
r\to\sqrt{l^2+r_0^2}+\delta r(t,l),\label{generalperturbations}
\end{equation}
which fixes the throat size $r_0$. A similar situation was considered in \cite{Bronnikov:2011if}, where various black holes supported by a nonminimal phantom scalar field were considered. There it was shown that, for the general spherically symmetric perturbations of the considered wormhole, (\ref{perturbations}) and (\ref{generalperturbations}), the gauge invariant quantity,
\begin{equation}\label{gaugeinvariant}
\chi(t,l)\propto r\delta\phi(t,l)-\frac{r^2}{l}\phi'(l)\delta r(t,l),
\end{equation}
satisfies the the wavelike equation (\ref{wavelike}) with the finite effective potential (\ref{potential}). When $\chi$ is finite at the throat, $\delta\phi$ does not diverge unless we fix the throat size by choosing $\delta r=0$ in~(\ref{gaugeinvariant}). Thus, the instability of the wormhole is easily understood, if one assumes perturbation of the throat's radius.

Summarizing all of the above, here we have shown that when imposing the correct boundary conditions for the perturbation, the Kanti-Kleihaus-Kunz wormholes proved to be unstable for whatever small values of the Gauss-Bonnet coupling constant $\alpha$. The instability develops after a long phase of damped quasinormal oscillations, which is similar to the time-domain profile of the eikonal instability observed for black holes in Einstein-Gauss-Bonnet theory. This may motivate further investigation of the possibility of nonhyperbolicity of the master wave equation for the wormhole case. The instability is driven by the purely imaginary mode which is nonperturbative in $\alpha$: that is, this mode does not go over into any finite mode in the limit $\alpha\to0$.

In other words the dynamical problem of evolution of the finite wave function $\chi$ which we solved for the finite effective potential~(\ref{potential}) does not have any divergence problem. The behavior of the dilaton field at the throat, as it was shown for the other type of a scalar field in \cite{Bronnikov:2011if}, is a pure artifact of the gauge chosen in \cite{Kanti:2011yv} and, therefore, can be safely ignored.

The spherically symmetric wormholes in the Gauss-Bonnet theory allow for the consistent dual-null formulation of the initial conditions for the nonlinear dynamics for massless matter. In particular, it was recently shown in~\cite{Shinkai:2017xkx} that the fate of a perturbed spherically symmetric wormhole supported by scalar fields is either a black hole or an expanding throat depending on the total energy of the structure. This result supports our general conclusion that the Kanti-Kleihaus-Kunz wormholes are unstable.

Even being unstable, the Kanti-Kleihaus-Kunz wormhole solution is a unique and important example of a traversable wormhole supported not by an exotic matter, but by introducing the second order curvature correction and dilaton, which are inspired by string theory. Thus, this wormhole appears naturally as a result of quantum corrections to the Einsteinian theory. Therefore, in our opinion, further efforts for finding stabilizing factors could be made in the future by considering nonminimal theory with additional fields (axions, fermions, gauge fields) or higher curvature corrections.

\acknowledgments{
M.~A.~C.~was supported by Coordena\c{c}\~ao de Aperfei\c{c}oamento de Pessoal de N\'ivel Superior (CAPES).
The publication has been prepared with the support of the ``RUDN University Program 5-100''.
A.~Z.~was supported by Conselho Nacional de Desenvolvimento Cient\'ifico e Tecnol\'ogico (CNPq), Brazil.}


\begin{thebibliography}{80}
\bibitem{Kanti:2011jz}
  P.~Kanti, B.~Kleihaus and J.~Kunz,
  Phys.\ Rev.\ Lett.\  {\bf 107}, 271101 (2011)
  [arXiv:1108.3003 [gr-qc]].

\bibitem{Tsukamoto:2017edq}
  N.~Tsukamoto,
  Phys.\ Rev.\ D {\bf 95}, no. 8, 084021 (2017)
  [arXiv:1701.09169 [gr-qc]].

\bibitem{Tsukamoto:2016qro}
  N.~Tsukamoto,
  Phys.\ Rev.\ D {\bf 94}, no. 12, 124001 (2016)
  [arXiv:1607.07022 [gr-qc]].

\bibitem{Konoplya:2005et}
  R.~A.~Konoplya and C.~Molina,
  Phys.\ Rev.\ D {\bf 71}, 124009 (2005)
  [gr-qc/0504139].

\bibitem{Konoplya:2010kv}
  R.~A.~Konoplya and A.~Zhidenko,
  Phys.\ Rev.\ D {\bf 81}, 124036 (2010)
  [arXiv:1004.1284 [hep-th]].

\bibitem{Aneesh:2018hlp}
  S.~Aneesh, S.~Bose and S.~Kar,
  Phys.\ Rev.\ D {\bf 97}, no. 12, 124004 (2018)
  [arXiv:1803.10204 [gr-qc]].

\bibitem{Zhou:2016koy}
  M.~Zhou, A.~Cardenas-Avendano, C.~Bambi, B.~Kleihaus and J.~Kunz,
  Phys.\ Rev.\ D {\bf 94}, no. 2, 024036 (2016)
  [arXiv:1603.07448 [gr-qc]].

\bibitem{Bueno:2017hyj}
  P.~Bueno, P.~A.~Cano, F.~Goelen, T.~Hertog and B.~Vercnocke,
  Phys.\ Rev.\ D {\bf 97}, no. 2, 024040 (2018)
  [arXiv:1711.00391 [gr-qc]].

\bibitem{Nascimento:2017def}
  J.~R.~Nascimento, A.~Y.~Petrov and P.~J.~Porfírio,
  arXiv:1712.08444 [gr-qc].

\bibitem{Nandi:2016uzg}
  K.~K.~Nandi, R.~N.~Izmailov, A.~A.~Yanbekov and A.~A.~Shayakhmetov,
  Phys.\ Rev.\ D {\bf 95}, no. 10, 104011 (2017)
  [arXiv:1611.03479 [gr-qc]].

\bibitem{Damour:2007ap}
  T.~Damour and S.~N.~Solodukhin,
  Phys.\ Rev.\ D {\bf 76}, 024016 (2007)
  [arXiv:0704.2667 [gr-qc]].

\bibitem{Konoplya:2016hmd}
  R.~A.~Konoplya and A.~Zhidenko,
  JCAP {\bf 1612}, no. 12, 043 (2016)
  [arXiv:1606.00517 [gr-qc]].

\bibitem{Gross:1986mw}
  D.~J.~Gross and J.~H.~Sloan,
  Nucl.\ Phys.\ B {\bf 291}, 41 (1987).

\bibitem{Metsaev:1987zx}
  R.~R.~Metsaev and A.~A.~Tseytlin,
  Nucl.\ Phys.\ B {\bf 293}, 385 (1987).

\bibitem{Kokkotas:2017ymc}
  K.~D.~Kokkotas, R.~A.~Konoplya and A.~Zhidenko,
  Phys.\ Rev.\ D {\bf 96}, no. 6, 064004 (2017)
  [arXiv:1706.07460 [gr-qc]].

\bibitem{Antoniou:2017hxj}
  G.~Antoniou, A.~Bakopoulos and P.~Kanti,
  Phys.\ Rev.\ D {\bf 97}, no. 8, 084037 (2018)
  [arXiv:1711.07431 [hep-th]].

\bibitem{Hartmann:2013tca}
  B.~Hartmann, J.~Riedel and R.~Suciu,
  Phys.\ Lett.\ B {\bf 726}, 906 (2013)
  [arXiv:1308.3391 [gr-qc]].

\bibitem{Doneva:2017bvd}
  D.~D.~Doneva and S.~S.~Yazadjiev,
  Phys.\ Rev.\ Lett.\  {\bf 120}, no. 13, 131103 (2018)
  [arXiv:1711.01187 [gr-qc]].

\bibitem{Silva:2017uqg}
  H.~O.~Silva, J.~Sakstein, L.~Gualtieri, T.~P.~Sotiriou and E.~Berti,
  Phys.\ Rev.\ Lett.\  {\bf 120}, no. 13, 131104 (2018)
  [arXiv:1711.02080 [gr-qc]].

\bibitem{Blazquez-Salcedo:2015ets}
  J.~L.~Bl\'azquez-Salcedo, L.~M.~Gonz\'alez-Romero, J.~Kunz, S.~Mojica and F.~Navarro-L\'erida,
  Phys.\ Rev.\ D {\bf 93}, no. 2, 024052 (2016)
  [arXiv:1511.03960 [gr-qc]].

\bibitem{Zhang:2017unx}
  H.~Zhang, M.~Zhou, C.~Bambi, B.~Kleihaus, J.~Kunz and E.~Radu,
  Phys.\ Rev.\ D {\bf 95}, no. 10, 104043 (2017)
  [arXiv:1704.04426 [gr-qc]].

\bibitem{Gwak:2017fea}
  B.~Gwak and D.~Ro,
  Eur.\ Phys.\ J.\ C {\bf 77}, no. 8, 554 (2017)
  [arXiv:1701.07737 [gr-qc]].

\bibitem{Blazquez-Salcedo:2016yka}
  J.~L.~Bl\'azquez-Salcedo {\it et al.},
  IAU Symp.\ {\bf 12} (S324), 265-272 (2016)
  [arXiv:1610.09214 [gr-qc]].

\bibitem{Antoniou:2017acq}
  G.~Antoniou, A.~Bakopoulos and P.~Kanti,
  Phys.\ Rev.\ Lett.\  {\bf 120}, no. 13, 131102 (2018)
  [arXiv:1711.03390 [hep-th]].

\bibitem{Nampalliwar:2018iru}
  S.~Nampalliwar, C.~Bambi, K.~Kokkotas and R.~Konoplya,
  Phys.\ Lett.\ B {\bf 781}, 626 (2018)
[arXiv:1803.10819 [gr-qc]].

\bibitem{Poisson:1995sv}
  E.~Poisson and M.~Visser,
  Phys.\ Rev.\ D {\bf 52}, 7318 (1995)
  [gr-qc/9506083].

\bibitem{Lobo:2005yv}
  F.~S.~N.~Lobo,
  Phys.\ Rev.\ D {\bf 71}, 124022 (2005)
  [gr-qc/0506001].

\bibitem{Eiroa:2003wp}
  E.~F.~Eiroa and G.~E.~Romero,
  Gen.\ Rel.\ Grav.\  {\bf 36}, 651 (2004)
  [gr-qc/0303093].

\bibitem{Dzhunushaliev:2017syc}
  V.~Dzhunushaliev, V.~Folomeev, B.~Kleihaus and J.~Kunz,
  Phys.\ Rev.\ D {\bf 97}, no. 2, 024002 (2018)
  [arXiv:1710.01884 [gr-qc]].

\bibitem{Bronnikov:2013coa}
  K.~A.~Bronnikov, L.~N.~Lipatova, I.~D.~Novikov and A.~A.~Shatskiy,
  Grav.\ Cosmol.\  {\bf 19}, 269 (2013)
  [arXiv:1312.6929 [gr-qc]].

\bibitem{Gonzalez:2008wd}
  J.~A.~Gonzalez, F.~S.~Guzman and O.~Sarbach,
  Class.\ Quant.\ Grav.\  {\bf 26}, 015010 (2009)
  [arXiv:0806.0608 [gr-qc]].

\bibitem{Bronnikov:2012ch}
  K.~A.~Bronnikov, R.~A.~Konoplya and A.~Zhidenko,
  Phys.\ Rev.\ D {\bf 86}, 024028 (2012)
  [arXiv:1205.2224 [gr-qc]].

\bibitem{Kanti:2011yv}
  P.~Kanti, B.~Kleihaus and J.~Kunz,
  Phys.\ Rev.\ D {\bf 85}, 044007 (2012)
  [arXiv:1111.4049 [hep-th]].

\bibitem{Takahashi:2010gz}
  T.~Takahashi and J.~Soda,
  Prog.\ Theor.\ Phys.\  {\bf 124}, 711 (2010)
  [arXiv:1008.1618 [gr-qc]].

\bibitem{Konoplya:2017lhs}
  R.~A.~Konoplya and A.~Zhidenko,
  JCAP {\bf 1705}, no. 05, 050 (2017)
  [arXiv:1705.01656 [hep-th]].

\bibitem{Reall:2014pwa}
  H.~Reall, N.~Tanahashi and B.~Way,
  Class.\ Quant.\ Grav.\  {\bf 31}, 205005 (2014)
  [arXiv:1406.3379 [hep-th]].

\bibitem{Konoplya:2011qq}
  R.~A.~Konoplya and A.~Zhidenko,
  Rev.\ Mod.\ Phys.\  {\bf 83}, 793 (2011)
  [arXiv:1102.4014 [gr-qc]].

\bibitem{Gundlach:1993tp}
  C.~Gundlach, R.~H.~Price and J.~Pullin,
  Phys.\ Rev.\ D {\bf 49}, 883 (1994)
  [gr-qc/9307009].

\bibitem{Bronnikov:2011if}
  K.~A.~Bronnikov, J.~C.~Fabris and A.~Zhidenko,
  Eur.\ Phys.\ J.\ C {\bf 71}, 1791 (2011)
  [arXiv:1109.6576 [gr-qc]].

\bibitem{Konoplya:2008ix}
  R.~A.~Konoplya and A.~Zhidenko,
  Phys.\ Rev.\ D {\bf 77}, 104004 (2008)
  [arXiv:0802.0267 [hep-th]].

\bibitem{Cuyubamba:2016cug}
  M.~A.~Cuyubamba, R.~A.~Konoplya and A.~Zhidenko,
  Phys.\ Rev.\ D {\bf 93}, no. 10, 104053 (2016)
  [arXiv:1604.03604 [gr-qc]].

\bibitem{Konoplya:2017ymp}
  R.~A.~Konoplya and A.~Zhidenko,
  Phys.\ Rev.\ D {\bf 95}, no. 10, 104005 (2017)
  [arXiv:1701.01652 [hep-th]].

\bibitem{Grozdanov:2016vgg}
  S.~Grozdanov, N.~Kaplis and A.~O.~Starinets,
  JHEP {\bf 1607}, 151 (2016)
  [arXiv:1605.02173 [hep-th]].

\bibitem{Grozdanov:2016fkt}
  S.~Grozdanov and A.~O.~Starinets,
  JHEP {\bf 1703}, 166 (2017)
  [arXiv:1611.07053 [hep-th]].

\bibitem{Konoplya:2017zwo}
  R.~A.~Konoplya and A.~Zhidenko,
  JHEP {\bf 1709}, 139 (2017)
  [arXiv:1705.07732 [hep-th]].

\bibitem{Gonzalez:2017gwa}
  P.~A.~Gonz\'alez, R.~A.~Konoplya and Y.~V\'asquez,
  Phys.\ Rev.\ D {\bf 95}, no. 12, 124012 (2017)
  [arXiv:1703.06215 [gr-qc]].

\bibitem{Shinkai:2017xkx}
  H.~A.~Shinkai and T.~Torii,
  Phys.\ Rev.\ D {\bf 96}, no. 4, 044009 (2017)
  [arXiv:1706.02070 [gr-qc]].
\end{thebibliography}
\end{document}